\documentclass[12pt,titlepage]{article}
\usepackage{graphicx}

\def\beq{\begin{equation}}
\def\eeq{\end{equation}}
\def\bea{\begin{eqnarray}}
\def\eea{\end{eqnarray}}

\title{Where is the $PdV$ term in the first law of black hole
thermodynamics?}

\author{{Brian P. Dolan}\\
{\small Department of Mathematical Physics, National University
of Ireland,}\\
{\small  Maynooth, Ireland}\\
{\small and}\\
{\small Dublin Institute for Advanced Studies,
10 Burlington Rd., Dublin, Ireland}\\
{\small e-mail: bdolan@thphys.nuim.ie}}

\begin{document}

\enlargethispage{-3mm}

\maketitle

\begin{abstract}
Traditional treatments of the first law of black hole thermodynamics
do not include a discussion of pressure and volume.
We give an overview of recent developments proposing a definition
of volume that can be used to extend the first law to include these
appropriately.  
New results are also presented relating to the
critical point and the associated second order phase transition
for a rotating black-hole in four-dimensional space-time which is
asymptotically anti-de Sitter.
In line with known results for a non-rotating charged black-hole,
this phase transition is shown to be
of Van der Waals type with mean field exponents.

\rightline{\small DIAS-STP-12-07}
\end{abstract}

\section{Introduction}

Ever since Hawking's discovery in 1974, \cite{Hawking1,Hawking2,Hawking3}, that black holes have a temperature associated to them, in the simplest case a temperature inversely proportional to their mass,
\beq \label{HawkingTemperature} T=\frac{\hbar}{8 \pi G M}
\eeq						
(we use units in which $c=1$),				
the thermodynamics of black holes has been a fascinating area of research.  
Equation (\ref{HawkingTemperature}) immediately implies that a Schwarzschild black hole in isolation is unstable: it will radiate and in so doing loses energy
hence the mass decreases,
thus increasing the temperature causing it to radiate with more power leading to a runaway effect.  

Hawking's result is fundamentally quantum mechanical in nature and came after a number of important developments in the classical thermodynamics of black holes. Penrose \cite {Penrose} realised that the mass of a rotating black hole can decrease, when rotational energy is extracted, and this  was followed by the observation that the area never decreases in any classical process.  
Nevertheless there is still a minimum, irreducible, mass below which one
cannot go classically \cite{Christodoulou,ChristodoulouRuffini}.
This lead Bekenstein's  to propose that an entropy should be associated with a black hole that is proportional to the area, $A$, of the event horizon, \cite{Bekenstein1,Bekenstein2} and Hawking's result (\ref{HawkingTemperature}) fixes the co-efficient 
to be one-quarter.  In natural units
\beq S=\frac 1 4 \frac {A} {\hbar G}, \eeq
where $\hbar G$ is the Planck length squared.

The first law of black hole thermodynamics, in its simplest form, associates the internal energy of a black hole with the mass, $U(S)=M$, (more precisely the ADM mass, as defined with reference to the time-like Killing vector at infinity
\cite{Wald}) and reads
\beq dU=TdS.
\eeq

The black hole instability referred to above is reflected in the
thermodynamic potentials by the fact that the heat capacity of a Schwarzschild black hole, 
\beq C=T\frac{\partial S}{\partial T}=-\frac{\hbar G}{8 \pi T^2}<0,\eeq
is negative.



The first law generalises to electrically charged, rotating black holes as
\beq \label{FirstLaw} d U=T d S+\Omega d J +\Phi d Q
\eeq							
where $J$ is the angular momentum of the black hole, $\Omega$ its angular velocity, 
and $Q$ the electric charge and  the electrostatic potential (see e.g. \cite{Wald}).

In contrast to elementary treatments of the first law of black hole thermodynamics it is noteworthy that (\ref{FirstLaw}) lacks the familiar $PdV$ term and there has recently been a number of papers addressing this 
issue, \cite{KP,KRT,TianWu,Dolan1,CGKP,Dolan2,Dolan3,KubiznakMann}.
This paper gives an overview of these developments and 
presents some new results in this regard, relating to the
presence of a critical point and a second order phase transition
for a rotating black-hole in four-dimensional asymptotically anti-de Sitter
space-time, a critical exponent that
is of a Van der Waals type with mean field exponents.

 A little thought shows that it is by no means obvious how to define the volume of a black hole.  For a Schwarzschild black hole the radial co-ordinate, $r$, is time-like inside the event horizon, where $r<r_h$, so it would 
seem non-sensical to associate a volume 
$V=4\pi\int_0^{r_h} r^2 d r=\frac{4\pi}{3} r_h^3$ with the black hole.  In fact identifying any function of $r_h$ alone with a volume, $V(r_h)$, will lead to inconsistencies in a thermodynamic description since the area, and hence the entropy, is already a function of $r_h$, $S=\pi r_h^2$, so any volume $V(r_h)$ would be determined purely in terms of the entropy.  
The internal energy, $U(S,V)$, should be a function of two variables, so
giving $V(S)$ uniquely as a specific function of $S$ is liable to lead to 
inconsistencies.
We shall see below how this potential problem is avoided.

\section{Pressure and Enthalpy}

From the point of view of Einstein's equations a pressure is associated with a cosmological constant. There is now very strong evidence that the cosmological constant in our Universe is positive \cite{Riess,Perlmutter}.  This poses a problem for the study of black hole thermodynamics for two reasons: firstly there is no asymptotic regime  in de Sitter space which allows the unambiguous identification of the ADM mass of a black hole embedded in a space with a positive $\Lambda$; secondly positive $\Lambda$ corresponds to negative pressure, implying thermodynamic instability.  The first problem is related to the fact that there are two event horizons for a de Sitter black hole, a black hole horizon and a cosmological horizon, and the radial co-ordinate is time-like for large enough values of $r$, outside the cosmological horizon. The second problem is not necessarily too serious as one can still glean some  information from negative pressure systems which are
thermodynamically unstable \cite{NegativeP}
(instability is not an insurmountable barrier to obtaining physical
information from a thermodynamic system,
after all, as described above, Hawking's formula (\ref{HawkingTemperature}) shows that black holes can have negative heat capacity but it is still a central formula in the understanding of black hole thermodynamics).  
In contrast for negative $\Lambda$  there is no cosmological horizon and the pressure is positive, the thermodynamics is perfectly well defined, so
we shall restrict our considerations here to negative $\Lambda$ and identify
the thermodynamic pressure $P=-\frac{\Lambda}{8\pi G}$ with the fluid dynamical pressure appearing in Einstein's equations.

The notion that the cosmological constant should be thought of as 
a thermodynamic variable is not new, and its thermodynamic conjugate
is often denoted $\Theta$ in the literature, \cite{TeitelboimHenneaux1,TeitelboimHenneaux2,
TeitelboimHenneaux3,Teitelboim,Sekiwa,Larranaga,Wangetal,Wang,Urano,LPPVP}, 
but $\Theta$ was not given a physical interpretation in these works.  

It may seem a little surprising to elevate $\Lambda$ to the status of a thermodynamic variable.  $\Lambda$ is usually thought of as a coupling constant in the Einstein action, on the same footing as Newton's constant, and it might seem 
bizarre to think of
Newton's constant as a thermodynamic variable.  However the nature of $\Lambda$ has long been mysterious \cite{Weinberg} and we should keep an open mind as to
its physical interpretation.
Indeed in \cite{KRT} it was argued that $\Lambda$ must be included in the
pantheon of thermodynamic variables for consistency with the Smarr relation
\cite{Smarr}, which is essentially dimensional analysis applied to thermodynamic
functions.  Furthermore \cite{KRT} suggested that, for a black hole embedded
in anti-de Sitter (AdS) space-time, the black hole mass is more
correctly interpreted as the enthalpy, $H$ beloved of chemists, rather than the
more traditional internal energy,
\beq \label{EnthalpyU}
M=H(S,P)=U(S,V)+ P V.
\eeq  
The $PV$ term in this equation can be though of as the contribution to the mass-energy of the 
black hole due the negative
energy density of the vacuum, $\epsilon = -P$, associated with a negative cosmological constant.
If the black hole has volume $V$ then it contains energy $\epsilon V=-PV$ and so the total
energy is $U=M-PV$.

This interpretation forces us to face up to
the definition of the black hole volume.  In \cite{KRT} $V$ is defined as the volume
relative to that of empty AdS space-time: the black hole volume is the volume 
that is excluded from empty AdS
when the black hole is introduced. We shall refer to this as the \lq\lq geometric
volume'' below. Other suggestions for the volume of a black hole have been made in
\cite{Parikh,BallikLake}

An alternative definition of the black-volume is that it is the thermodynamic
conjugate of the pressure, under the Legendre transform (\ref{EnthalpyU}),
\beq \label{ThermodynamicV}
V:=\frac{\partial H}{\partial P}, \eeq
which we shall call the \lq\lq thermodynamic volume''.

With the definition of the thermodynamic volume (\ref{ThermodynamicV})
we are in a position to state the definitive version
of the first law of black hole thermodynamics,
\beq \label{CompleteFirstLaw}
{d U = T d S+\Omega d J +\Phi d Q - PdV}
\eeq
which follows from the Legendre transform of
\beq\label{dH}
d M = dH = T d S+\Omega d J +\Phi d Q + V d P.
\eeq
Equation (\ref{dH}), in $\Theta \,d \Lambda$ notation, appeared 
in \cite{CCK}.  A $PdV$ term was considered in \cite{Pad} and \cite{TianWu},
but only for $J=0$, when $V$ and $S$ cannot be considered to be independent
thermodynamic variables.

\section{Thermodynamic volume}

The suggested definition of the thermodynamic volume (\ref{ThermodynamicV})
must be tested for consistency.  For example, for a non-rotating black hole
in four-dimensional space-time, the line element is given, in Schwarzschild
co-ordinates, by,
\beq
d^2s = -f(r) dt^2 + f^{-1}(r)dr^2 + r^2 d\Omega^2, \eeq
with 
\beq \label{fdef}
f(r) = 1 -\frac{2 m}{r} - \frac{\Lambda}{3} r^2, 
\eeq
and $d\Omega^2=d\theta^2+\sin^2\theta d\phi^2$ the
solid angle area element.\footnote{From now on we set $G=\hbar = 1$
to avoid cluttering formulae.}  The event horizon is defined by $f(r_h)=0$,
\beq \label{rh}
\frac \Lambda 3 r_h^3 - r_h + 2m=0,
\eeq
but we do not need to solve this equation explicitly in order to analyse 
(\ref{ThermodynamicV}).  We already know that 
\beq\label{SandP} 
S={\pi r_h^2}, \qquad P=- \frac{\Lambda}{8\pi}
\eeq
and, for negative $\Lambda$, the ADM mass is $M=m$ \cite{TeitelboimHenneaux3}, 
which, following the philosophy of \cite{KRT}, we identify with the enthalpy, $H(S,P)$.  
Solving (\ref{rh}) for $m$ immediately yields 
\beq m= \frac {r_h}{2} \left(1  - \frac{\Lambda}{3} r_h^2 \right),
\label{mass}
\eeq
from which $H(S,P)=M=m$, with (\ref{SandP}), identifies the enthalpy as
\beq \label{EnthalpySchwaarzschild}
H(S,P)=\frac {1} {2} \left(\frac {S}{\pi}\right)^{\frac 1 2} 
\left(1+\frac {8 S P} {3} \right).\eeq

The usual thermodynamic relations can now be used to determine the temperature
and the volume,
\bea \label{TH}
T&=&\left(\frac{\partial H}{\partial S} \right)_P\quad\Rightarrow\quad
T=\frac {1} {4}\left(\frac{1}{\pi S }\right)^{\frac 1 2}
(1+8   P S)=\frac{ (1-\Lambda r_h^2)}
{4\pi r_h}\;, \\
\label{VH}
V&=& \left(\frac{\partial H}{\partial P} \right)_S\quad
\Rightarrow\quad
V= \frac 4 3 \frac{S^{\frac 3 2}}{\sqrt{\pi}} = \frac {4 \pi r_h^3}{3}.
\eea 
That the resulting thermodynamic volume (for a non-rotating black hole) is identical to the
geometric volume is quite remarkable, but appears co-incidental as this
equality no longer holds for rotating (Kerr-AdS) black holes, as we shall see.
It does however hold for non-rotating black holes
in all dimensions \cite{Pad,TianWu,Dolan1}.

As mentioned in the introduction, equation (\ref{VH}) has a potential problem
associated with it, in that it implies that the volume and the entropy
cannot be considered to be independent thermodynamic variables, $S$
determines $V$ uniquely -- they cannot be varied independently and so
$V$ seems redundant.  Indeed this may the reason
why $V$ was never considered in the early literature on black hole thermodynamics.  But this is an artifact of the non-rotating approximation, $V$ and $S$
can, and should, be considered to be independent variables for a rotating
black hole.

The line element for a charged rotating black hole in 4-dimensional AdS
space is \cite{Carter}
\beq  \label{ChargedAdSKerr}
d s^2=-\frac{\Delta}{\rho^2}\left( d t - \frac{a\sin^2\theta}{\Xi}\, d\phi \right)^2
+\frac{\rho^2}{\Delta}dr^2 + \frac{\rho^2}{\Delta_\theta}d\theta^2 
+\frac{\Delta_\theta\sin^2\theta}{\rho^2}\left(a d t - \frac{r^2 + a^2}{\Xi}\,d\phi \right)^2,
\eeq
where 
\bea \label{MetricFunctions}
\Delta&=&\frac{(r^2+a^2)(L^2+r^2)}{L^2} -2 m r + q^2, \qquad
\Delta_\theta=1-\frac{a^2}{L^2}\cos^2\theta,\nonumber \\
\rho^2& = & r^2 + a^2\cos^2\theta, \qquad \Xi = 1-\frac{a^2}{L^2},
\eea
and the cosmological constant is $\Lambda=-\frac 3 {L^2}=-8\pi P$.

The physical properties of this space-time are well known 
\cite{TeitelboimHenneaux3}.
The metric parameters $m$ and $q$ are related to the ADM mass $M$ and the 
electric charge $Q$ by
\beq \label{MassCharge}
M=\frac{m}{\Xi^2},\qquad Q=\frac {q} {\Xi}.\eeq
The event horizon, $r_+$, lies at the largest root of $\Delta(r)=0$, so,
in terms of geometrical parameters,
\beq \label{Mass}
M=\frac{(r_+^2+a^2)(L^2+r_+^2) + q^2 L^2}{2 r_+ L^2 \,\Xi^2} \eeq
and the area of the event horizon is
\beq \label{Area}
A=4\pi\frac{r_+^2+a^2}{\Xi},
\eeq
giving
\beq 
S=\pi\frac{r_+^2+a^2}{\Xi}.
\eeq

The angular momentum is $J=a M$ and the relevant thermodynamic
angular velocity is
\beq
\Omega=\frac{a(L^2+r_+^2)}{L^2(r_+^2+a^2)}.  
 \eeq
As explained in \cite{CCK}, $\Omega$ here is the difference between the
asymptotic angular velocity and the angular velocity at the black hole 
outer horizon.

The electrostatic potential, again the difference between the
potential at infinity and at the horizon, is
\beq
\Phi=\frac{q r_+}{r_+^2+a^2}.
\eeq

To determine the thermodynamic properties, $M$ must be expressed in terms of $S$, $J$, $Q$
and $P$ (or, equivalently, $L$).  This was done in \cite{CCK} and the result is

\beq \label{Enthalpy}
H(S,P,J,Q):=
\frac {1}{2}\sqrt{\frac{
\left(S   + \pi Q^2 + \frac{8 P S^2}{3}\right)^2 + 
4 \pi^2\left( 1+\frac{8 P S}{3}  \right) J^2}
{\pi S}}.
\eeq
This generalises the Christodoulou-Ruffini formula \cite{Christodoulou,ChristodoulouRuffini} for the mass
of a rotating black hole in terms of its irreducible mass, $M_{irr}$.  
(The irreducible mass for a black hole with entropy $S$
is the mass of a Schwarzschild black hole with the same 
entropy, $M^2_{irr}=\frac {S}{4\pi}$).

The temperature follows from
\bea \label{CCKTemperature}
T&=& \left.\frac {\partial H}{\partial S}\right|_{J,Q,P}\\
&=&\frac {1}{8\pi H}\left[
\left(1  +\frac{\pi Q^2}{S}+\frac {8 P S}{3} \right)
\left(1 - \frac{\pi Q^2}{S}+ 8 P S \right)
-4\pi^2 \left(\frac {J}{S}\right)^2\right],\nonumber
\eea
from which we immediately see that $T\ge 0$ requires
\beq
J^2\le
\frac {S^2}{4\pi ^2}\left(1  +\frac{\pi Q^2}{S}+\frac {8 P S}{3} \right)
\left(1 - \frac{\pi Q^2}{S}+ 8 P S \right).
\eeq
The maximum angular momentum,
\beq \label{Jmax}
|J_{max}|=
\frac{S}{2\pi}\sqrt{\left(1  +\frac{\pi Q^2}{S}+\frac {8 P S}{3} \right)
\left(1 - \frac{\pi Q^2}{S}+ 8 P S \right)},
\eeq
is associated with an extremal black hole.

From (\ref{ThermodynamicV}) and (\ref{Enthalpy}) the thermodynamic volume is \cite{Dolan2}
\beq \label{VolumeSQPJ}
V= \left.\frac {\partial H}{\partial P}\right|_{S,J,Q}=\frac{2}{3 \pi H}\left[S\left(S + \pi Q^2 + \frac{8PS^2}{3}\right)+ 2\pi^2 J^2  \right],
\eeq
which is manifestly positive.  

The angular velocity and the electric potential also follow from (\ref{Enthalpy}) via
\beq \label{ThermodynamicOmega}
\Omega=\left.\frac{\partial H}{\partial J}\right|_{S,Q,P}=
 \frac {4\pi^2 J\left(1+\frac{8PS}{3} \right)}{2 H \sqrt{\pi S}}
\eeq
and
\beq \label{ThermodynamicPhi}
\Phi=\left.\frac{\partial H}{\partial Q}\right|_{S,J,P}=
 \frac {2\pi Q\left(S+\pi Q^2 +\frac{8PS^2}{3} \right)}{2 H \sqrt{\pi S}}.
\eeq
The Smarr relation follows from (\ref{Enthalpy}), (\ref{CCKTemperature}),
(\ref{VolumeSQPJ}), (\ref{ThermodynamicOmega}) and (\ref{ThermodynamicPhi}), namely
\beq
\frac{H}{2} +  PV -ST - J\Omega -\frac {Q\Phi}{2}=0,
\eeq
from which it is clear that the $PV$-term must be included for consistency,
as pointed out in \cite{KRT}.

It is clear from (\ref{VolumeSQPJ}) that,
in general, $V$ is a function of all 
the four independent thermodynamical variables, $S$, $P$, $J$ and $Q$,
but for the limiting case $J=0$,
\beq
V=\frac{4}{3} \frac{ S^{\frac 3 2 }}{\sqrt\pi},
\eeq
is determined purely in terms of $S$ alone, independent of both  $P$ and $Q$.
Thus, as explained in the introduction, $V$ and $S$ cannot be viewed
as thermodynamically independent variables as $J\rightarrow 0$, rendering the description in terms of the thermodynamic potential $U(S,J)$ impossible
in this limit.

Expressing the thermodynamic volume (\ref{VolumeSQPJ}) in terms of geometrical variables one gets \cite{Dolan2}
\beq \label{Volume}
V = \frac {2\pi }{3}\left\{
\frac{(r_+^2+a^2)\bigl(2 r_+^2 L^2 + a^2 L^2 - r_+^2 a^2 \bigr) +  L^2 q^2 a^2}
{L^2 \Xi^2 \,r_+}\right\}.
\eeq
Given that the area of the event horizon is
\beq A = 4\pi\frac{r_+^2 + a^2}{\Xi}
\eeq
then, if we define a na\"ive volume 
\beq \label{IsoPerimetric}
V_0 := \frac{r_+ A}{3} = 
\frac {4\pi}{3}\frac{r_+(r_+^2 + a^2)}{\Xi},
\eeq
equations (\ref{Mass}) and (\ref{Volume}) give
\beq
V= V_0 + \frac {4\pi a^2 M} {3}  = V_0 + \frac{4\pi} 3 \frac {J^2}{M},
\eeq
a formula first derived in \cite{CGKP}.  As pointed out in that reference,
equation (\ref{IsoPerimetric}) implies that the surface to volume
ratio of a black hole is always less than that of a sphere with radius $r_+$
in
Euclidean geometry.  This is the opposite of our usual intuition
that a sphere has the smallest surface to volume ratio of any closed
surface --- the isoperimetric inequality of Euclidean geometry.
Thus the surface to volume ratio of a black hole satisfies the reverse
of the usual isoperimetric inequality
(a similar result holds in higher dimensions \cite{CGKP}).  At least this seems to be the case
if quantum gravity effects are not taken into account.  In one case
where quantum gravity corrections can be calculated using the techniques
in \cite{MaloneyWitten} , the three-dimensional Ba\~nados--Zanelli--Teitelboim
(BTZ) black hole \cite{BTZ}, they
tend to reduce the black hole volume \cite{Dolan1} so it seems possible 
that quantum gravity effects may affect the reverse isoperimetric inequality.

\section{The First Law}

To examine the consequences of the $PdV$ term in the first law
we need to perform a Legendre transform on the enthalpy to
obtain the internal energy $U(S,V,J,Q)$ from $U=H-PV$. 
We first write
the enthalpy (\ref{Enthalpy}) in the form
\beq H=\sqrt{a + b P +c P^2},
\eeq
where
\bea 
a&:=& \frac{\pi}{S}\left\{ \frac 1 4 \left( \frac S \pi + Q^2 \right)^2 + J^2 \right\} \nonumber \\
b &:=& \frac {4\pi}{3}\left\{ \frac S \pi \left( \frac S \pi + Q^2 \right)+2J^2 \right\}
\\
c &:=& \left(\frac{4\pi}{3}\right)^2\left(\frac{S}{\pi}\right)^3.\nonumber 
\eea
Note that the discriminant,
\beq b^2 - 4ac = \frac{64 \pi^2}{9} J^2 
\left(J^2 +\frac {S Q^2}{\pi}\right), 
\eeq
is positive.

Now
\beq V=\left.\frac {\partial H}{\partial P}\right|_{S,J,Q}
= \frac{b + 2 c P}{2 H}
\qquad \Rightarrow \qquad P=\frac{2 H V - b}{2 c}.
\eeq
This allows us to re-express $H$ as a function of $V$,
\beq H=\frac 1 2 \sqrt{\frac {b^2 - 4ac}
{V^2 - c}}.
\eeq
We can immediately conclude that 
\beq
V^2\ge c=\left(\frac{4\pi}{3}\right)^2\left(\frac{S}{\pi}\right)^3,
\eeq
with equality only when
\beq b^2 - 4ac = \frac{64 \pi^2}{9} J^2 
\left(J^2 +\frac {S Q^2}{\pi}\right)=0,
\eeq
{\it i.e.} when $J=0$.

It is now straightforward to determine
\beq
U=H-PV= H- \left(\frac {H V^2}{c} - \frac{b V}{2c}\right)
=  \frac{b V}{2c} - 
\frac {\sqrt{\bigl(V^2 - c\bigr) \bigl(b^2-4ac\bigr)}}{2c},
\eeq
which immediately gives
\bea \label{InternalEnergy}
U(S,V,J,Q)&=&  \left(\frac {\pi}{S}\right)^3\left[ 
\left(\frac{3V}{4\pi}\right)
\left\{ \left(\frac S {2\pi}\right)\left(\frac S {\pi} + Q^2 
\right)+J^2 \right\}\vline  height 20pt  width 0pt depth 20pt
\right.\\
& & \kern 50pt \left. -|J|\left\{\left(\frac{3V}{4\pi}\right)^2-\left(\frac S \pi \right)^3\right\}^{\frac 1 2}
\left(\frac{S Q^2}{\pi}+J^2\right)^{\frac 1 2}
 \right].  \nonumber
\eea

Note the subtlety in the $J\rightarrow 0$ limit, (\ref{InternalEnergy})
is not differentiable at $J=0$ unless
\beq
\Bigl(\frac{3V}{4\pi}\Bigr)^2=\Bigl(\frac S \pi \Bigr)^3
\eeq
there.

Equation (\ref{InternalEnergy}) can now be used to study the efficiency
of a Penrose process.
If a black hole has initial mass $M_i$, with internal energy $U_i$,
and is taken through a quasi-static series of thermodynamic steps to a state
with final internal energy $U_f$, then energy can be extracted if $U_f<U_i$.
This is the thermodynamic description of a Penrose process \cite{Penrose}
and the efficiency is
\beq \eta=\frac{U_i-U_f}{M_i}.
\eeq

We can determine the maximum efficiency for a process at constant $P$
by
first expressing $U$ in (\ref{InternalEnergy}) in terms of $S$, $P$, $J$
and $Q$:
\beq \label{USPJQ}
U= \frac{\Bigl( S   + \pi Q^2\Bigr)
\left(S   + \pi Q^2 + \frac{8 P S^2 }{3} \right)
+ 4\pi^2 \left(1 + \frac{4 P S} {3} \right) J^2 }
{2\sqrt{ \pi S \left[
\left(S   + \pi Q^2 + \frac{8 P S^2}{3}\right)^2 + 
4 \pi^2\left( 1+\frac{8 P S}{3}  \right) 
J^2\right]}},
\eeq
which is manifestly positive.

For simplicity consider first the $Q=0$ case, for which
\beq dU = T d S +\Omega dJ- P d V. 
\eeq
The work extracted at any infinitesimal step is 
\beq \label{dW}
dW=-dU=-TdS -\Omega dJ + P dV
\eeq
and, since $dS\ge 0$, this is maximised in an isentropic process $dS=0$.
Now with $Q=0$ and $S$ and $P$ held constant, the internal energy
in equation (\ref{USPJQ}) can be thought
of as a function of $J$ only, $U(J)$.  The greatest efficiency is then
obtained by starting with an extremal black hole and
reducing the angular momentum from
$J_{max}$ to zero,
it is given by
\beq \eta_{ext} = \frac{U(J_{max})-U(0)}{H(J_{max})}
\eeq
where $H(J_{max})=M_{ext}$ is the initial extremal mass.
One finds
\beq \eta_{ext}=\frac{1+2 P S} {1+4PS}-
\frac {1}{\sqrt{2+8PS}}\frac {3} {\left(3+8PS\right)}.
\eeq
In asymptotically flat space, $\Lambda=0$, we set $P=0$ in $\eta_{ext}$
and obtain the famous result \cite{Wald}
\beq \eta_{ext}=1 - \frac{1}{\sqrt{2}}\approx 0.2929.
\eeq
More generally, $\eta_{ext}$ is a maximum for 
$SP=1.837\ldots$ (obtained by solving
a quartic equation) and attains there the value $0.5184\ldots\;$.
Thus turning on a negative cosmological constant increases the efficiency
of a Penrose process, as first observed in \cite{Dolan2}.

What is happening here is that, as $|J|$ decreases (giving a positive contribution
to $d W$) the volume decreases, which actually tends to {\it decrease}
the work done because of the $P d V$ term in  (\ref{dW}).  
But when $P>0$, the extremal value $|J_{max}|$ in (\ref{Jmax})
is increased, which more than compensates, and overall $\eta_{ext}$ is increased.

For a charged black hole the internal energy is a function of
$J$ and $Q$ for an isobaric isentropic process, $U(J,Q)$.  The requirement
$J^2_{max}\ge 0$ in (\ref{Jmax}) imposes the constraint
\beq Q^2\le Q_{max}^2 = \left(\frac{S}{\pi}\right)\left(1+ 8 P S \right)\eeq
on the charge.  The greatest efficiency is achieved starting from an extremal
black hole with $Q^2=Q_{max}^2$ and reducing both $J$ and $Q$ to zero in the
final state,
\beq \eta_{ext}=\frac{U(J_{max},Q_{max})-J(0,0)}{H(J_{max},Q_{max})}=
\frac{3}{2}\left(\frac{1+8PS}{3+16 PS}\right) ,\eeq
with ${H(J_{max},Q_{max}})$ the initial extremal mass, $M_{ext}$.
For large $S$ efficiencies of up to 75\% are possible \cite{Dolan2}, 
which should
be compared to 50\% in the $\Lambda=0$ case, \cite{Wald}.

\section{Critical Behaviour}

With knowledge of both $H$ and $U$ general questions concerning the heat capacity of
black holes can be addressed.
The heat capacity at constant volume, 
$C_V=T/
\left(\frac {\partial T} {\partial S} \right)_{V,J,Q}$, 
tends to zero when $J=0$, though $C_V$ can be non-zero for $J\ne 0$
it does not diverge. 
For comparison the heat capacity at constant pressure, 
$C_P=T/
\left(\frac {\partial T} {\partial S} \right)_{P,J,Q}$,
$C_P$ vanishes when $T=0$ and diverges when 
$\frac {\partial T} {\partial S}$ =0. 

A full stability analysis was given in \cite{CCK} and there
are both local and global phase transitions. Local stability
can be explored visually, by plotting thermodynamic functions,
or analytically, examining the curvature of the derivatives of 
thermodynamics functions.

\subsection{$Q=0$}

Let us first focus on the $Q=0$ case.  The red (lower) curve in the figure below shows
the locus of points where $C_P$ diverges in the $J-S$ plane, it is given by 
setting the denominator of $C_P$,
\bea \label{CPdivergence}
&&144\,(\pi {J}{P})^{4} \left( 9+32\,SP \right) +24\,(\pi {P}{J})^{2}
({PS})^{2} \left(3+ 16\,SP \right)  \left( 3+8\,SP \right) ^
{2}\nonumber\\
&&\hskip 60pt
-({P}{S})^{4} \left(1- 8\,SP\right)  \left( 3+8\,SP \right)^{3},
\eea
to zero.
The green (upper) curve is the $T=0$ locus, all points above and left of this
curve are unphysical as $T<0$ in this region.

\bigskip

\begin{figure}[htb]
\centering
\includegraphics[width=90mm]{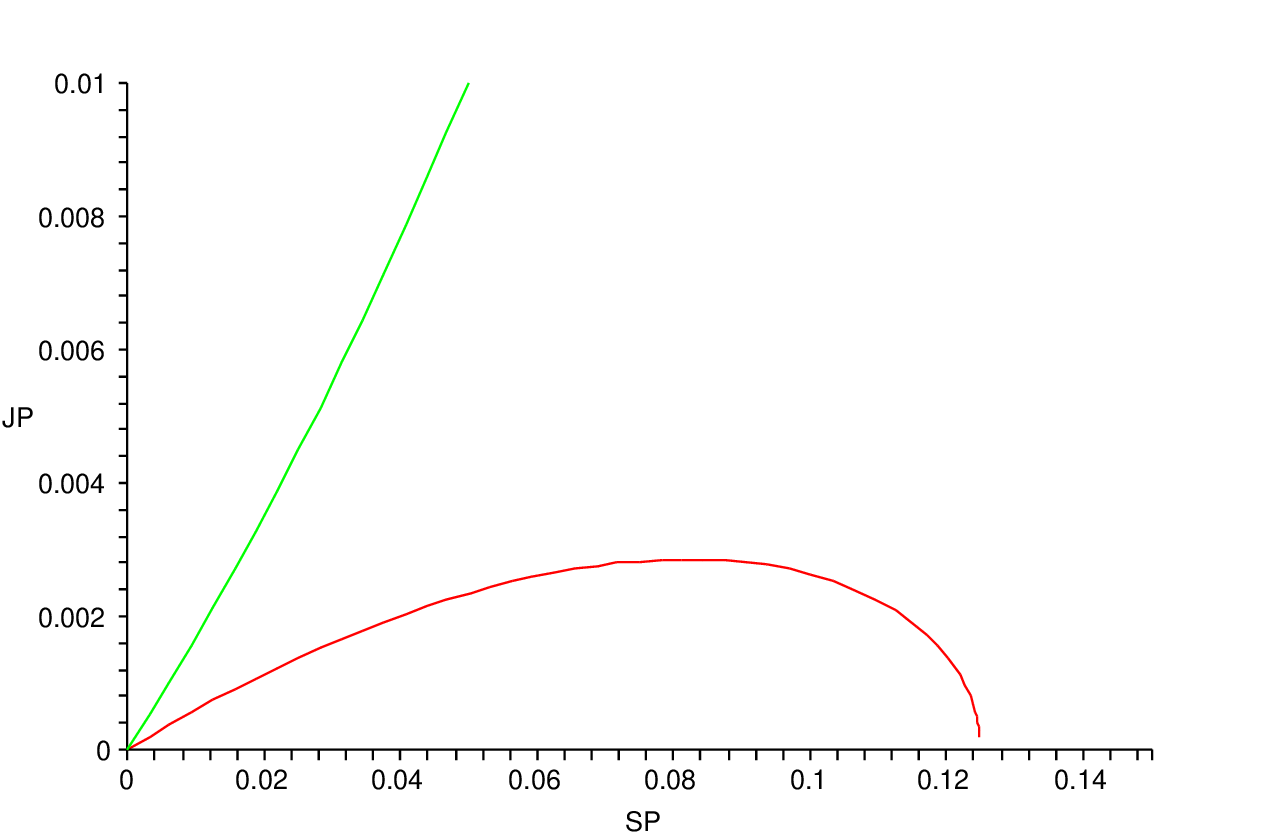}
\caption{$T=0$ and $C_P\rightarrow\infty$ curves in $J-S$ plane.}
\end{figure}

There is also a global phase transition, not shown in the figure, 
when the free energy of pure AdS is lower than that of a black hole
in asymptotically AdS space-time, the famous Hawking-Page phase 
transition \cite{HawkingPage}.
We shall focus on the second order local phase transition here and examine its critical properties.

In general, at fixed $P$ and $J$, there are two values of $S$ at which
$C_P$ diverges, and there is a critical point where these two values coalesce
into one, the maximum of the lower curve in figure 1.
This critical point was first identified in 
\cite{CCK}.  
On purely dimensional grounds $P C_P$ can be expressed as 
a function of $PS$ and $PJ$ and the critical point can be found
analytically, by solving
a cubic equation, but the explicit form is not very illuminating.
Numerically it lies at 

\beq (PS)_{crit}\approx 0.08204, \qquad (PJ)_{crit}\approx 0.002857.
\eeq
The critical temperature is obtained from (\ref{CCKTemperature}), with $Q=0$,
\beq
\left(\frac {T} {\sqrt{P}}\right)_{crit}\approx 0.7811
\eeq
and the critical volume likewise from (\ref{VolumeSQPJ})
\beq
\left(V P^{3/2}\right)_{crit}\approx 0.01768
\eeq
(the authors of \cite{CCK} fix 
$P=\frac {3}{8\pi}\approx 0.1194$, corresponding to $L=1$, and find a critical 
value of $J$ at $J_c\approx 0.0236$).

The equation of state cannot be obtained analytically, but its properties
near the critical point can be explored by a series expansion and critical
exponents extracted.
Define the reduced temperature and volume as
\beq t=\frac {T-T_c}{T_c} \qquad v=\frac {V-V_c}{V_c}.
\eeq
It is convenient to expand around the critical point using 
\beq
p:= 16\pi\Bigl(PJ - (PJ)_{crit}\Bigr)
\eeq
and
\beq
q:=8\Bigl(PS - (PS)_{crit}\Bigr).
\eeq
Expanding the temperature (\ref{CCKTemperature}) around the critical point,
with $Q$ set to zero, gives 
\beq \label{tps}
t = 2.881\, p + 2.201\, p q + 0.3436\, q^3 +o(p^2, p q^2,q^4).
\eeq
while similar expansion of the thermodynamic volume (\ref{VolumeSQPJ})
yields
\beq \label{vps}
v=-10.44 \, p + 2.284 \, q +o(p^2, p q,q^2).
\eeq
For a given fixed $J>0$, $p$ is the deviation from
critical pressure in units of $1/(16\pi J)$, but one must be aware
that this interpretation precludes taking the $J\rightarrow 0$ limit
in this formulation.  
Bearing this in mind,
(\ref{tps}) and (\ref{vps}) give 
the $J>0$, $Q=0$ equation of state parametrically in terms of $q$.
Eliminating $q$ one arrives at
\beq \label{EoS}
p=0.3472 \,t - 0.1161 \, t v - 0.02883 \,v^3 +o(t^2, t v^2,v^4).
\eeq

The critical exponent $\alpha$ is defined by
\beq C_V\propto t^{-\alpha}
\eeq
and, since as already stated, $C_V$ does not diverge at $t=0$, $\alpha=0$.
To see this explicitly note that 
$C_V=T/\frac{\partial T}{\partial S}\bigr|_V$ and 
\beq
\left.\frac{\partial T}{\partial S}\right|_V
=\left.\frac{\partial T}{\partial S}\right|_P
+\left.\frac{\partial T}{\partial P}\right|_S 
\left.\frac{\partial P}{\partial S}\right|_V
=T_c \left(\left.\frac{\partial t}{\partial S}\right|_P
+\left.\frac{\partial t}{\partial P}\right|_S 
\left.\frac{\partial P}{\partial S}\right|_V\right).
\eeq
Now, near the critical point, (\ref{tps}) gives
\bea
\left.\frac{\partial t}{\partial S}\right|_P&=&8 P
\left.\frac{\partial t}{\partial q}\right|_p = o(p,q^2),\\
\left.\frac{\partial t}{\partial P}\right|_S&=&8 S
\left.\frac{\partial t}{\partial q}\right|_p +16\pi J\left.\frac{\partial t}{\partial p}\right|_q = 2.881 (16\pi J)+ o(p,q),
\eea
while (\ref{vps}) implies $d p =0.2188\, d\, q$ for constant $v$, from
which is follows that
$\left.\frac{\partial P}{\partial S}\right|_V$
is non-zero at the critical point,
hence $\left.\frac{\partial T}{\partial S}\right|_V$ does not vanish
at the critical point and so 
$\alpha=0$.

The exponent $\beta$ is defined by
\beq v_> - v_< = |t|^\beta
\eeq
where $v_>$ is the greater volume and $v_<$ the lesser volume across
the phase transition, at constant pressure, when $t<0$
($v_<$ is negative, since $v=0$ at the critical point).

\begin{figure}[htb]
\centering
\includegraphics[height=80mm]{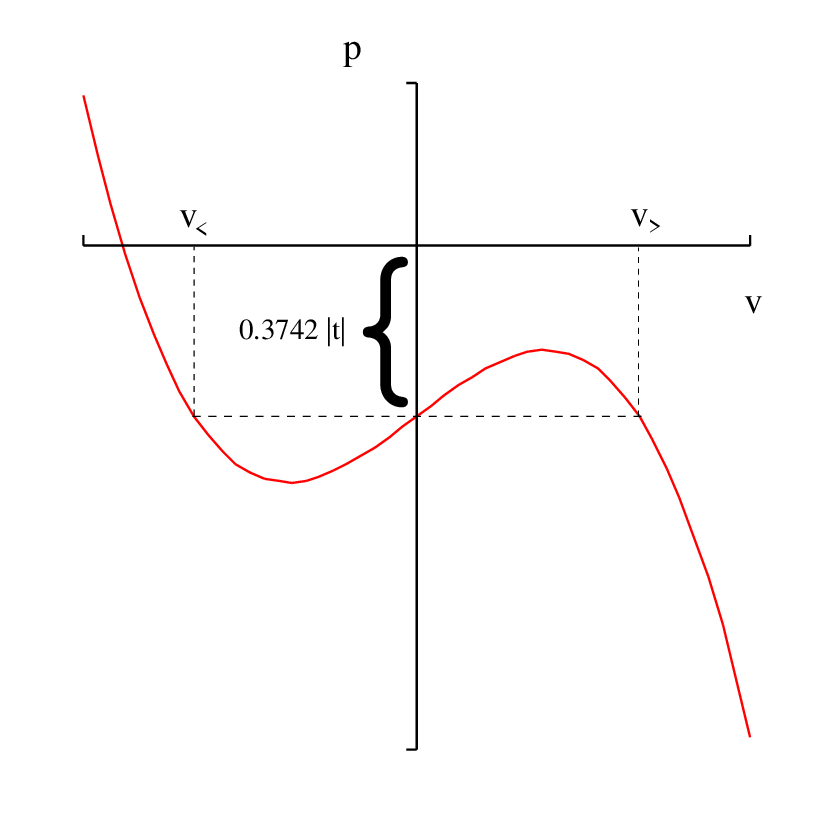}
\caption{Construction associated with Maxwell's equal area law.}
\end{figure}

Keeping $p$ and $t$ constant in (\ref{EoS}) 
implies that
\beq
p\int_{v_<}^{v_>}  dv =  0.3742\, t\int_{v_<}^{v_>} dv -
\int_{v_<}^{v_>}\Bigl(0.1161 \, t v + 0.02883 \,v^3 \Bigr)d v.
\eeq
Allowing for the area of the rectangle in figure 2, 
namely $0.3742\, |t|(v_> - v_<)$, Maxwell's equal area law then requires
\beq
\int_{v_<}^{v_>}\Bigl(0.1161 \, t v + 0.02883 \,v^3 \Bigr)d v =0
\qquad \Rightarrow \qquad
|t| \propto  (v_>^2 + v_<^2). 
\eeq
It is clear from the figure that $v_> - v_< \gg v_> + v_<$ so
\beq
(v_>^2 + v_<^2) = \frac 1 2 \bigl((v_> - v_<)^2 +  (v_> + v_<)^2\bigr)
\approx  \frac 1 2 (v_> - v_<)^2
\eeq
giving
\beq |t|\propto  (v_> - v_<)^2 \eeq
and $\beta =\frac 1 2 $.

The critical exponent $\gamma$ is related to the isothermal compressibility,
\beq
\kappa_T=-\frac {1} {V} \left( 
\frac{\partial V}{\partial P}\right)_{T,J}
=-\frac {1} {V} \left( 
\frac{\partial V}{\partial P}\right)_{S,J}
-\left(\frac{\partial V}{\partial S}\right)_{P,J}
\frac{\left(\frac{\partial T}{\partial P}\right)_{S,J}}
{\left(\frac{\partial T}{\partial S}\right)_{P,J}} 
\eeq
which diverges along the same curve as $C_P$ does
(the adiabatic compression, $\kappa_S=-\frac {1} {V} \left( 
\frac{\partial V}{\partial P}\right)_{S,J}$, is everywhere finite ---
see equation (\ref{kappaS})).
$\gamma$ gives the divergence of the isothermal compressibility
near the critical point,
\beq
\kappa_T\propto t^{-\gamma}.
\eeq
$\gamma$ can be found by expanding the denominator of $C_P$
in (\ref{CPdivergence})
around the critical point,
but a quicker method, since we have the equation of state, is to differentiate
(\ref{EoS}) with respect to $v$, keeping $t$ constant, giving
\beq
\left.\frac{\partial p}{\partial v}\right|_t\propto -t,
\eeq
hence 
\beq
\kappa_T\propto -\left.\frac{\partial v}{\partial p}\right|_t \propto \frac 1 t
\eeq
and $\gamma=1$.

Lastly  setting $t=0$ in (\ref{EoS}) we see that
\beq |p|\propto |v|^\delta
\eeq 
with $\delta=3$, again the mean field result. 

To summarise,  the critical exponents are
\beq
{\alpha=0,\quad \beta=\frac 1 2,\quad \gamma=1,\quad \delta=3.}
\eeq
These are the same critical exponents as the Van der Waals fluid
and, more importantly, are mean field exponents.\footnote{While this
manuscript was in preparation we became aware that these critical
exponents have also been found, using a virial expansion approximation,
in \cite{GKM}.}

It was first pointed in \cite{CEJM1,CEJM2}
that a non-rotating, charged black hole
has a critical point of the same nature as that of 
of a Van der Waals fluid, and the critical exponents for the black hole phase transition in this case were
calculated in \cite{KubiznakMann} and verified to be mean field exponents,
which are indeed the those of a Van der Waals fluid.
A similarity between the
neutral rotating black hole and the Van der Waals phase transition was
first pointed out in \cite{CCK} and further explored in \cite{Dolan2}.

The critical point can be visualised by plotting the Gibbs free energy
\beq G(T,P,J)=H(S,P,J)-TS, \eeq 
for $J=1$ and $Q=0$, as a function of $P$ and $T$ as in figure 3. 
We see the \lq\lq swallow-tail
catastrophe'' that is typical of the Van der Waals phase 
transition \cite{Sewell2}.

\begin{figure}[htb]
\centering
\includegraphics[height=110mm]{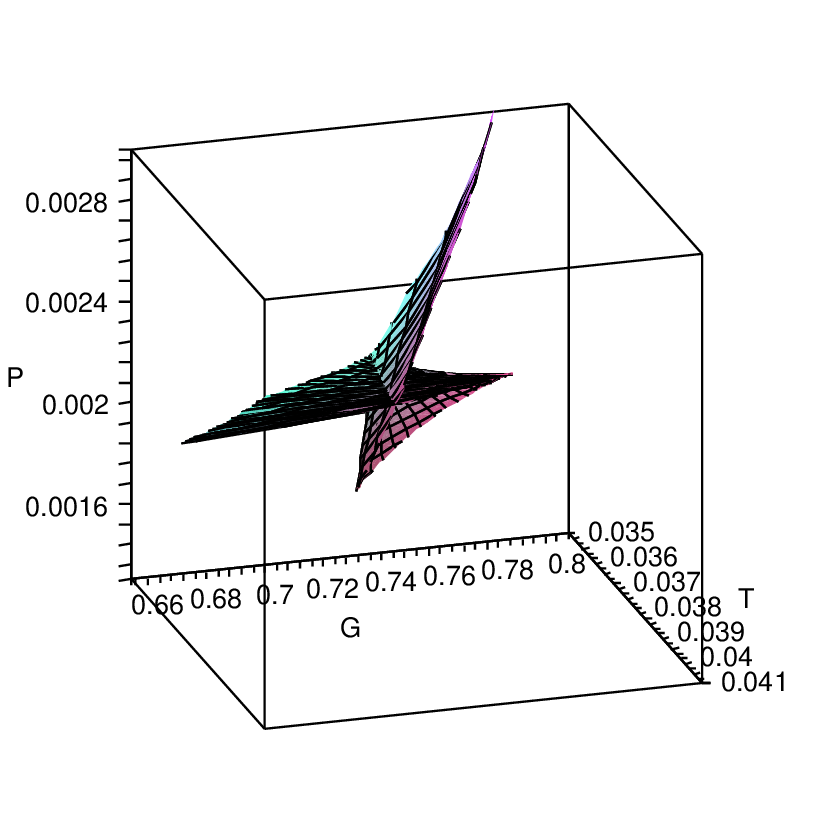}
\caption{Gibbs free energy as a function of pressure and temperature, at
fixed angular momentum.}
\end{figure}

This structure is a straightforward consequence of Landau theory, \cite{LL}.  
Near the critical point the Landau free energy is
\beq \label{Landau}
L(T,P,v)= G(T,P) + A\left\{(p-B t) v +C t v^2 + D v^4\right\}+\ldots\;,
\eeq
where $G(T,P)$ is the 
Gibbs free energy and $A$, $B$, $C$ and $D$ are positive constants
(for simplicity the constant $J$ is not made explicit).
As stressed in \cite{Goldenfeld} $L$ 
is not strictly speaking a thermodynamic function as
it depends on three variables, $p$, $t$ and $v$ instead of two:
$v$ is to be
determined in terms of $p$ and $t$ by extremising $L$ to obtain the
equation of state.

For notational convenience equation (\ref{Landau}) can be written,
for fixed $p$ and $t$, as
\beq \label{A4}
L  = a + b v + c v^2 + v^4
\eeq
where $a$, $b$ and $c$ need not be positive  and $L\rightarrow \frac 1 {AD} L$
has been rescaled by a trivial positive constant.
We are to think of $b$ and $c$ are control parameters that can be
varied by varying $p$ and $t$.

Extremising (\ref{A4}) with respect to $v$ determines the value
of $v$ in terms of $b$ and $c$ through
\beq \label{bextremum}
b= -2 c v - 4 v^3.
\eeq
Using this in $L$ leads to
\beq \label{Lextremum}
L = a-c v^2 - 3 v^4.
\eeq
Equations (\ref{bextremum}) and (\ref{Lextremum}) together give $L(a,b,c)$ implicitly: a parametric plot of $L(b,c)$,
for any fixed $a$, reveals a characteristic \lq\lq swallow-tail 
catastrophe'' structure.
With hindsight the swallow-tail structure is clear: in 
the $A-D-E$ classification of critical points of functions, \cite{Arnold1},
(\ref{A4}) has three control parameters and is derived from type $A_4$
in Arnold's classification.



\subsection{$Q\ne 0$}
The above structure was first found in AdS black hole thermodynamics in the charged $J=0$ case \cite{CEJM1,CEJM2}, where the equation of state can be found exactly and the critical exponents can be determined \cite{KubiznakMann}. 
When both $J$ and $Q$ are non-zero an analytic analysis
is much more difficult, for example finding the zero locus of the denominator
of $C_P$ requires solving a quintic equation.
However numerical studies show that for a charged rotating black hole, 
as long as the
charge is below the extremal value, the picture is
qualitatively the same: the critical exponents are the same, the Landau free
energy is still related to type $A_4$ and the Gibbs free energy still 
takes on a characteristic swallow-tail shape.  For fixed values of $J$
and $Q$, not both zero,  all that changes is the
numerical value of the co-efficients in equations (\ref{tps}), (\ref{vps})
and (\ref{EoS}) or, equivalently the numerical values of the constants
$A$, $B$, $C$ and $D$ in (\ref{Landau}). As long as none of these constants
actually changes sign the nature of the critical point does not change 
and the critical exponents are the same.

As first observed in \cite{CCK} there is a line of second order
critical points in the $J-Q$ plane, as shown below,

\begin{figure}[htb]
\centering
\includegraphics[height=80mm]{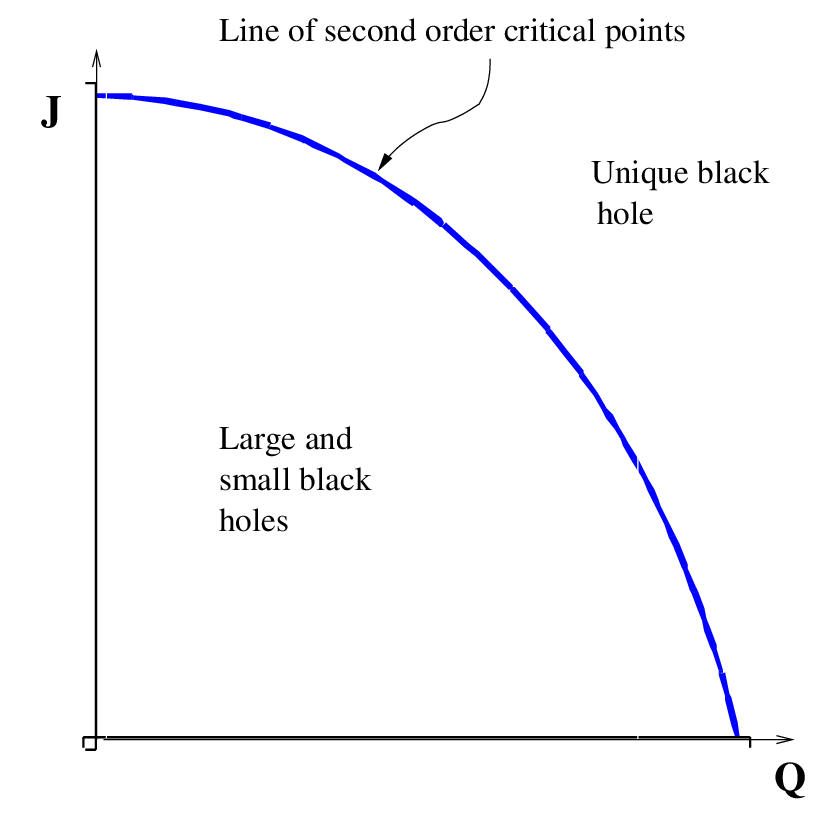}
\caption{The line of second order critical points in the $J-Q$ plane, \cite{CCK}.}
\end{figure}

As long as $J$ and $Q$ are not both zero conservation of charge/angular 
momentum protects the black hole against the Hawking-Page phase transition
to pure anti-de sitter.

\section{Compressibility and the speed of sound}

In the previous section, the nature of the singularity in the isothermal
compressibility near the critical point was discussed, but the adiabatic
compressibility
\beq
\kappa_S=-\frac {1} {V} \left( 
\frac{\partial V}{\partial P}\right)_{T,J,Q}
\eeq
is also of interest, and this was studied in \cite{Dolan3} 
on which most of this section is based.  
From (\ref{ThermodynamicV}) one finds, 
setting $Q=0$ for simplicity, that
\beq\label{kappaS}
\kappa_S = \frac{36(2\pi J)^4 S}
{(3+8PS)
\bigl\{  (3+8PS)S^2 +  (2\pi J)^2 \bigr\}
\bigl\{2(3+8PS)S^2 + 3(2\pi J)^2 \bigr\}
}.
\eeq
This is finite at the critical point, indeed it never diverges for
any finite values of $S$, $P$ and $J$, and it vanishes as $J\rightarrow 0$:
non-rotating black holes are completely incompressible.
Black holes are maximally compressible in the extremal case $T=0$,
when $J=J_{max}$ in (\ref{Jmax}),
\beq\label{Extremalbeta}
\left.\kappa_S\right|_{T=0}=
\frac {2\,S \left( 1+8 PS \right) ^{2}}
{\left( 3+8 PS\right)^2  \left( 1+4 PS\right)}.
\eeq

A speed of sound, $c_S$, can also be associated with the black hole, in the usual
thermodynamic sense that
\beq c_S^{-2} = \left.\frac  {\partial \rho}{\partial P}\right|_{S,J}= 1 + \rho \,\kappa_S 
=1+
\frac {9\,(2\pi J)^4}
{ \left\{ 2(3 + 8 PS)S^2 + 3\,(2\pi J)^2 \right\}^2}\;,
\eeq
where $\rho=\frac M V$ is the density.
$c_S$ is unity for incompressible non-rotating black holes
and is lowest for extremal black holes in which case
\beq \label{SoundSpeed}
\left.c_S^{-2}\right|_{T=0}=
1+\left(\frac{1+8 PS}{3+8 PS}\right)^2.
\eeq
giving $c^2_S=0.9$ (in units with $c=1$) when $P=0$.
In the limiting case $PS\rightarrow \infty$,  $c_S^2$ 
achieves a minimum value of $1/2 $.

These results show that the equation of state is very stiff 
for adiabatic variations of non-rotating black holes and 
gets softer as $J$ increases.  For comparison, the adiabatic compressibility of a degenerate gas of $N$ relativistic 
neutrons in a volume $V$ at zero temperature
follows from the degeneracy pressure
\beq
P_{deg}=(3\pi^2)^{\frac 1 3}\frac{c\hbar}{4}\left(\frac{V}{N}\right)^{-\frac 4 3 }\quad \Rightarrow \quad \kappa_S = \frac 3 {4P_{deg}}.\eeq
For a neutron star $\frac N V\approx 10^{45}\ m^{-3}$ and 
$\kappa_S\approx 10^{-34}\ kg^{-1} \,m \,s^2$.
With zero cosmological
constant the black hole adiabatic compressibility 
at zero temperature is given by (\ref{Extremalbeta}) with $P=0$,
\beq \label{betaZeroP}
\left.\kappa_S\right|_{T=P=0}= \frac {2S}{9}=\frac{4 \pi M^2 G^3}{9 \,c^8},\eeq
where the relevant factors of $c$ and $G$ are included, and the entropy has
been set to the extremum value of $2\pi M$.
Putting in the numbers
\beq\left.\kappa_S\right|_{T=0}=2.6\times 10^{-38}\left(\frac{M}{M_\odot}\right)^2 \,kg^{-1}\,m\,s^2,
\eeq
which is four orders of magnitude less than that of
a solar-mass neutron star.  We conclude that the 
zero temperature black hole equation of state, although \lq\lq softer''
than that of a non-rotating black hole, is still very much stiffer than that of
a neutron star.

The \lq\lq softest'' compressibility for a neutral black hole however is the isothermal compressibility: for an extremal black hole
\beq
\kappa_{T}\bigl|_{T=0}=\frac{2S\bigl(11+80PS+128(PS)^2\bigr)}{(1+4PS)\bigl(3+48PS+128(PS)^2\bigr)}
\quad\mathop{\longrightarrow}_{P\rightarrow 0}\quad\frac{22\,S}{3},
\eeq
some 33 times larger than $\kappa_{S}\bigr|_{T=P=0}$ in (\ref{betaZeroP}),
but still much larger than degenerate matter in a solar-mass neutron star.

\section{Discussion}

The obvious open question arising from the ideas presented here is:
what about $\Lambda >0$?

The analysis of critical behaviour in \S 5 is only valid for $\Lambda<0$,
this critical point lies deep in the region $P>0$ and does not appear
to be of any relevance to astrophysical situations.  It is certainly
of interest in the AdS-CFT correspondence \cite{AdSCFT}
but the particular analysis
of \S 5, being in $1+3$-dimensions could only be relevant to $2+1$-dimensional
conformal field theory, which is of course of interest in its own
right \cite{Hartnoll}.  One could perform a similar
analysis for $4+1$-dimensional, or yet higher dimensional black holes,  
to try and 
gain insight into higher dimensional conformal field theory, and indeed
this seems to have been the motivation in \cite{CEJM1,CEJM2,CGKP}, 
but the analysis of angular momentum in higher dimension is more
difficult and will be left for future work.

The thermodynamics of black holes in de Sitter space-time is a notoriously
difficult problem \cite{Sekiwa,Wang,Urano,GomberoffTeitelboim,DehghaniKhajehAzad,RCP,CorichiGomberoff,Aros} as there are two event horizons and no
\lq\lq asymptotically de Sitter'' region inside the cosmological horizon.
Even with no black hole, a na\"ive interpretation of the cosmological
horizon implies that
the transition from $\Lambda=0$ to any infinitesimally small $\Lambda>0$
appears to involve a discontinuous jump from zero to infinite entropy, at
least if one associates the usual Hawking-Bekenstein entropy with the cosmological horizon when $\Lambda>0$.   

Nevertheless it is argued in \cite{GomberoffTeitelboim} that a consistent
strategy is to fix the relevant components of the metric at the cosmological
horizon, rather than at spacial infinity as would be done in
asymptotically flat or AdS space-time.   When that is done the same
expression for the ADM mass (\ref{Mass}) is obtained, but with $L^2\rightarrow -L^2$, so $\Xi>1$ while the angular momentum is still given by $J=aM$.
In this picture, all of the formulae in \S3 are applicable for positive
$\Lambda$ and negative $P$, provided $P$ is not too negative.
If $\Lambda$ is too large the black hole horizon and the cosmological horizon
coincide and demanding that this does not happen puts a lower
bound on $P$, for any fixed $S$, $J$ and $Q$:
with $Q=0$, for example, this requirement constrains $P$ to
\beq
P>\frac{\sqrt{S^2 + 12\pi J^2} -2 S}{8S}.
\eeq
Provided $P$ lies above this lower bound
we can analytically continue (\ref{Enthalpy}) to negative $P$,
with the understanding that $S$ is the entropy of the black hole event
horizon only and does not include any contribution from the cosmological
horizon. 

Of course $P<0$ is thermodynamically unstable, but it can be argued, in some circumstances at least, that positive pressures can be analytically
continued to negative pressures {\cite{NegativeP,LL}, and in a cosmological
context there can now be little doubt that $P<0$.
Adopting the strategy of \cite{GomberoffTeitelboim} the maximal efficiency of a rotating
black hole in de Sitter space will be less than in the $\Lambda=0$ case,
based on simply changing the sign of $\Lambda$ in \S4, and the zero charge
efficiency
vanishes when the black hole horizon and the cosmological horizon coincide 
at $PS=-\frac 1 8$.
Any such deviation from the $\Lambda=0$ case will however
be completely negligible for astrophysical black holes around one solar mass and the observed value of $\Lambda$, but it could be more significant during
periods of inflation when $\Lambda$ was larger.

It has been suggested that primordial black-holes may have formed in the
early Universe \cite{Carr} and, if this is the case and if they formed in sufficient
numbers at any stage, then one should model the primordial gas as containing
a distribution of highly incompressible black holes, like beads in a gas.
These would certainly be expected to affect the overall compressibility
of the gas as well as the speed of sound through the gas. In a radiation 
dominated Universe, ignoring the matter density, the speed of sound
in the photon gas would be given by
\beq
c_\gamma^{-2}=\left.\frac{\partial \epsilon}{\partial P}\right|_S = 3 c^{-2},
\eeq
where $\epsilon$ is the energy density
(essentially since the equation of state is $P=\frac 1 3 \epsilon$)
so $c_\gamma=0.577 \,c$. Since the speed of sound associated with the embedded
black hole \lq\lq beads'' is $c_S\ge \sqrt{0.9}\,c= 0.9487\,c$ the presence of a
significant density of primordial black holes would expected to affect
speed of sound in the photon gas and thus affect the dynamics.

\begin{figure}[htb]
\centering
\includegraphics[height=80mm]{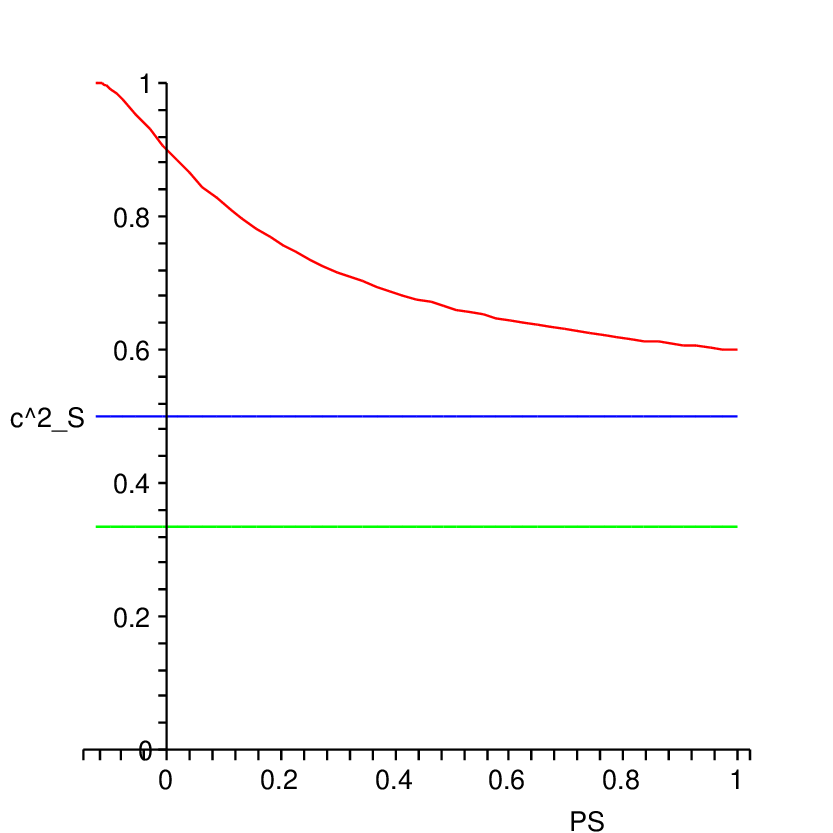}
\caption{The speed of sound for an electrically neutral, extremal, black hole
(with $c=1$).}
\end{figure}

The square of the speed of sound for an extremal
electrically neutral black hole is plotted in figure 5, for $PS>-1/8$. For comparison
the asymptotic value ($c^2_S=1/2$ for $PS\rightarrow \infty$) and the speed
of sound in a thermal gas of photons ($c_\gamma^2=1/3$) are also shown.

\section{Conclusions}

In conclusion there are strong reasons to believe that the cosmological
constant should be included in the laws of black hole thermodynamics as
a thermodynamic variable, proportional to the pressure of ordinary thermodynamics.
The conjugate variable is a thermodynamic volume (\ref{ThermodynamicV}) and
the complete first law of black hole thermodynamics is now (\ref{FirstLaw}),
\beq 
{d U = T d S+\Omega d J +\Phi d Q - PdV}.
\eeq

With this interpretation the ADM mass of the black hole is identified 
with the enthalpy 
\beq
M=H(S,P,J,Q) = U(S,V,J,Q) + PV
\eeq
rather than the internal energy, $U$,  of the system.

The inclusion of this extra term increases the maximal efficiency
of a Penrose process: for a neutral black hole 
in asymptotically anti-de Sitter space 
the maximal efficiency is increased from $0.2929$ in asymptotically
flat space to 0.5184 in the asymptotically AdS case. For a charged black hole the efficiency
can be as high as 75\%.  A positive cosmological constant is expected to reduce
the efficiency of a Penrose process below the asymptotically flat space
value.

This point of view makes the relation between asymptotically AdS black holes
and the Van der Waals gas, first found in \cite{CEJM1,CEJM2}, even closer
as there is now a critical volume associated with the critical point.
The thermodynamic volume then plays the r\^ole of an order parameter for
this phase transition and the critical exponents take the mean field values,
\beq
{\alpha=0,\quad \beta=\frac 1 2,\quad \gamma=1,\quad \delta=3.}
\eeq

While there is no second order phase transition for a black hole in de Sitter space, there are other possible physical effects of including the $PdV$ term in the first law.   The adiabatic compressibility can be calculated 
(\ref{Extremalbeta}) and
the speed of sound for such a black hole (\ref{SoundSpeed}) is greater even than that of a photon gas and approaches $c$ when $PS=-1/8$. 

Despite much progress the thermodynamics of black holes in de Sitter space-time is still very poorly understood and no doubt much still remains to be discovered.

\bibliographystyle{nar}
\bibliography{Van-der-Waals-1.1}

\begin{thebibliography}{10}

\bibitem{Hawking1}
Hawking, S.~W. (1974)
{\em Nature} {\bf 248}, 30.

\bibitem{Hawking2}
Hawking, S.~W. (1975)
{\em Comm. Math. Phys.} {\bf 43}, 199
Erratum: ibid. vol. 46, p. 206 (1976).

\bibitem{Hawking3}
Hawking, S.~W. (1976)
{\em Phys. Rev.} {\bf D13}, 191.

\bibitem{Penrose}
Penrose, R. (1969)
{\em Riv. Nuovo Cimento} {\bf 1}, 252.

\bibitem{Christodoulou}
Christodoulou, D. (1970)
{\em Phys. Rev. Lett.} {\bf 25}, 1596.

\bibitem{ChristodoulouRuffini}
Christodoulou, D. and Ruffini, R. (1971)
{\em Phys. Rev.} {\bf D4}, 3552.

\bibitem{Bekenstein1}
Bekenstein, J.~D. (1972)
{\em Lett. Nuovo. Cimento} {\bf 4}, 737.

\bibitem{Bekenstein2}
Bekenstein, J.~D. (1973)
{\em Phys. Rev.} {\bf D7}, 2333.

\bibitem{Wald}
Wald, R.~M. (1984)
General Relativity,
University of Chicago Press, .

\bibitem{KP}
Kothawala, D. and Padmanabhan, T. (2009)
{\em Phys Rev} {\bf D79}, 104020
[arXiv:0904.0215 [gr-qc]].

\bibitem{KRT}
Kastor, D., Ray, S., and Traschen, J. (2009)
{\em Class. Quantum Grav.} {\bf 26}, 195011
[arXiv:0904.2765 [hep-th]].

\bibitem{TianWu}
Tian, Y. and Wu, X. (2011)
{\em Phys Rev} {\bf D83}, 021501
[arxiv:1007.4331].

\bibitem{Dolan1}
Dolan, B.~P. (2011)
{\em Class. Quantum Grav.} {\bf 28}, 125020
[arXiv:1008.5023].

\bibitem{CGKP}
Cvetic, M., Gibbons, G.~W., Kubiz\u{n}\'{a}k, D., and Pope, C.~N. (2011)
{\em Phys. Rev.} {\bf D84}, 024037
[arXiv:1012.2888[hep-th].

\bibitem{Dolan2}
Dolan, B.~P. (2011)
{\em Class. Quantum Grav.} {\bf 28}, 235017
[arXiv:1106.6260].

\bibitem{Dolan3}
Dolan, B.~P. (2011)
{\em Phys. Rev.} {\bf D84}, 127503
[arXiv:1109.0198].

\bibitem{KubiznakMann}
Kubiz\u{n}\'{a}k, D. and Mann, R.~B. (2012)
{\em JHEP} {\bf 1207}, 033
[arXiv:1205.0559].

\bibitem{Riess}
Riess, A.~G. et al. (1998)
{\em Astronomical Journal} {\bf 116}, 1009.

\bibitem{Perlmutter}
Perlmutter, S. et al. (1999)
{\em Astrophysical Journal} {\bf 517}, 565
[arXiv:astro-ph/9812133].

\bibitem{NegativeP}
Luk\'acs, B. and Martin\'as, K. (1990)
{\em Acta. Phys. Pol.} {\bf B21}, 177.

\bibitem{TeitelboimHenneaux1}
Henneaux, M. and Teitelboim, C. (1984)
{\em Phys.~Lett.} {\bf 143B}, 415.

\bibitem{TeitelboimHenneaux2}
Henneaux, M. and Teitelboim, C. (1989)
{\em Phys.~Lett.} {\bf 222B}, 195.

\bibitem{TeitelboimHenneaux3}
Henneaux, M. and Teitelboim, C. (1985)
{\em Commun. Math. Phys.} {\bf 98}, 391.

\bibitem{Teitelboim}
Teitelboim, C. (1985)
{\em Phys.~Lett.} {\bf 158B}, 293.

\bibitem{Sekiwa}
Sekiwa, Y. (2006)
{\em Phys. Rev.} {\bf D73}, 084009
[arXiv:hep-th/0602269].

\bibitem{Larranaga}
{Larra{\~n}aga Rubio}, E.~A.
Stringy generalization of the first law of thermodynamics for rotating {BTZ}
  black hole with a cosmological constant as state parameter
[arXiv:0711.0012 [gr-qc]] (2007).

\bibitem{Wangetal}
Wang, S., Wu, S., Xie, F., and Dan, L. (2006)
{\em Chin. Phys. Lett.} {\bf 23}, 1096
[arXiv:hep-th/0601147].

\bibitem{Wang}
Wang, S.
Thermodynamics of high dimensional {Schwarzschild} {de Sitter} spacetimes:
  variable cosmological constant
[arXiv:gr-qc/0606109] (2006).

\bibitem{Urano}
Urano, M., Tomimatsu, A., and Saida, H. (2009)
{\em Class. Quantum Grav.} {\bf 26}, 105010
arXiv:0903.4230 [gr-qc].

\bibitem{LPPVP}
Lu, H., Pang, Y., Pope, C.~N., and {Vazquez-Poritz}, J. (1212)
{\em Phys. Rev.} {\bf D86}, 044011
[arxiv:1204.1062].

\bibitem{Weinberg}
Weinberg, S. (1989)
{\em Rev. Mod. Phys.} {\bf 61}, 1.

\bibitem{Smarr}
Smarr, L. (1973)
{\em Phys. Rev. Lett.} {\bf 30}, 71
Erratum: ibid. vol. 30, p. 521 (1973).

\bibitem{Parikh}
Parikh, M.~K. (2006)
{\em Phys. Rev.} {\bf D73}, 124021.

\bibitem{BallikLake}
Ballik, W. and Lake, K.
The volume of stationary black holes and the meaning of the surface gravity
[arXiv:1005.1116 [gr-qc]] (2012).

\bibitem{CCK}
Caldarelli, M.~M., Cognola, G., and Klemm, D. (2000)
{\em Class. Quantum Grav.} {\bf 17}, 399
[arXiv:hep-th/9908022].

\bibitem{Pad}
Padmanabhan, T. (2002)
{\em Class. Quantum Grav.} {\bf 19}, 5387
[arXiv:gr-qc/0204019].

\bibitem{Carter}
Carter, B. (1968)
{\em Comm. Math. Phys.} {\bf 10}, 280.

\bibitem{MaloneyWitten}
Maloney, A. and Witten, E. (2010)
{\em JHEP} {\bf 1002}, 029
[arXiv:0712.0155 [hep-th]].

\bibitem{BTZ}
Ba{\~n}ados, M., Teitelboim, C., and Zanelli, J. (1992)
{\em Phys. Rev. Lett.} {\bf 69}, 1849.

\bibitem{HawkingPage}
Hawking, S.~W. and Page, D.~N. (1983)
{\em Comm. Math. Phys.} {\bf 87}, 577.

\bibitem{GKM}
Gunasekaran, S., Kubiz\u{n}\'{a}k, D., and Mann, R.~B.
Extended phase space thermodynamics for charged rotating black holes and
  born-infeld vacuum polarization
[arXiv:1208.6251].

\bibitem{CEJM1}
Chamblin, A., Emparan, R., Johnson, C.~V., and Myers, R.~C. (1999)
{\em Phys. Rev.} {\bf D60}, 064018
[arXiv:hep-th/9902170v2].

\bibitem{CEJM2}
Chamblin, A., Emparan, R., Johnson, C.~V., and Myers, R.~C. (1999)
{\em Phys. Rev.} {\bf D60}, 104026
[arXiv:hep-th/9904197].

\bibitem{Sewell2}
Sewell, M.~J. (1977)
{\em Math. Proc. Camb. Phil. Soc.} {\bf 82}, 147.

\bibitem{LL}
Landau, L.~D. and Lifshitz, E.~M. (1980)
Course of Theoretical Physics, Part 1: Statistical Physics, volume {\bf 5},
Pergammon, Oxford,  3rd edition
p44.

\bibitem{Goldenfeld}
Goldenfeld, N. (1992)
Lectures On Phase Transitions And The Renormalization Group (Frontiers in
  Physics; 85),
Addison-Wesley, .

\bibitem{Arnold1}
Arnold, V.~I. (1984)
Catastrophe Theory,
Springer-Verlag, .

\bibitem{AdSCFT}
Aharony, O., Gubser, S.~S., Maldacena, J., Ooguri, H., and Oz, Y. (2000)
{\em Phys. Rep.} {\bf 323}, 183
[arXiv:hep-th/9905111].

\bibitem{Hartnoll}
Hartnoll, S.~A. (2009)
{\em Class. Quant. Grav.} {\bf 26}, 224002
[arXiv:0903.3246v3 [hep-th]].

\bibitem{GomberoffTeitelboim}
Gomberoff, A. and Teitelboim, C. (2003)
{\em Phys. Rev.} {\bf D67}, 104024
[arXiv:hep-th/0302204].

\bibitem{DehghaniKhajehAzad}
Dehghani, M.~H. and KhajehAzad, H. (2003)
{\em Can. J. Phys.} {\bf 81}, 1363
[hep-th/0209203].

\bibitem{RCP}
{Roy Choudhury}, T. and Padmanabhan, T. (2007)
{\em Gen. Rel. Grav.} {\bf 39}, 1789
[arXiv:gr-qc/0404091].

\bibitem{CorichiGomberoff}
Corichi, A. and Gomberoff, A. (2004)
{\em Phys. Rev.} {\bf D69}, 064016
[arXiv:hep-th/0311030].

\bibitem{Aros}
Aros, R. (2008)
{\em Phys. Rev.} {\bf D77}, 104013
[arXiv:0801.4591].

\bibitem{Carr}
Carr, B.~J. (2003)
{\em Lect. Notes Phys.} {\bf 631}, 301
[arXiv:astro-ph/0310838].

\end{thebibliography}

\end{document}